\documentstyle[aps,prl,multicol,epsf]{revtex}
\begin{document}
\title{
{\small Submitted to Phys. Rev. B, 30 July 1998} \vskip 5mm
Quantum Monte Carlo calculation of Compton profiles of solid lithium}

\author{Claudia Filippi$^{\;*}$ and David M. Ceperley$^{\;*\dag}$}
\address{Department of Physics$^{\;*}$ and National Center for Supercomputing 
Applications$^{\;\dag}$, \\University of Illinois at Urbana-Champaign, Urbana, 
Illinois 61801}
\maketitle
\begin{abstract}
Recent high resolution Compton scattering experiments in lithium have 
shown significant discrepancies with conventional band theoretical results.
We present a pseudopotential quantum Monte Carlo study of electron-electron 
and electron-ion correlation effects on the momentum distribution of lithium.
We compute the correlation correction to the valence Compton profiles 
obtained within Kohn-Sham density functional theory in the local density
approximation and determine that electronic correlation does not account 
for the discrepancy with the experimental results. 
Our calculations lead do different conclusions than recent GW studies and 
indicate that other effects (thermal disorder, core-valence separation etc.) 
must be invoked to explain the discrepancy with experiments.
\end{abstract}

\begin{multicols}{2}
\setcounter{collectmore}{5}

\section{Introduction}
\label{s1}

Inelastic x-ray scattering is called Compton scattering when the energy and 
the momentum transferred are large compared to the characteristic energy and 
reciprocal interelectronic distance of the scattering system.
Compton scattering probes the electronic structure of materials through the
electronic momentum distribution and,
if the scattering system is metallic, gives direct
information on various characteristics of the Fermi surface, such as position 
and size of the Fermi breaks and their renormalization due to electron-electron 
correlation.
Fermi momenta are of the order of 1 a.u. so that high-resolution in 
momentum are necessary in order to resolve features related to the Fermi
surface. In the last few years, the advent of high intensity, high 
energy and well-polarized synchrotron sources has made possible to obtain 
resolutions of the order of 0.1 a.u. and high statistics. Record resolutions
of the order of 0.02 a.u. have been recently achieved for solid 
beryllium~\cite{HamalainenBe} and lithium~\cite{HamalainenLi}.

In the range of energy and momentum transferred where the recoil 
electron can be considered free, {\it i.e.}
within the impulse approximation~\cite{Eisenberger1}, the experimental 
double-differential Compton cross section is related to the momentum 
distribution $n({\bf p})$ of the electronic system as (in atomic units 
$\hbar=e=m=1$):
\begin{eqnarray}
\frac{{\rm d}^2\sigma}{{\rm d}{\rm \Omega}\,{\rm d}{\rm \omega}_2}&=&
\left(\frac{{\rm d}\sigma}{{\rm d}{\rm \Omega}}\right)_{\mbox{\tiny\rm Th}}
\,\times\nonumber\\& &
\int\frac{{\rm d}\,{\bf p}}{(2\pi)^3}\;n({\bf p})\;
\delta\left(q^2/2+{\bf p}\cdot{\bf q}-\omega\right),
\end{eqnarray}
where $\omega$ and ${\bf q}$ are the energy and momentum transferred,
$\omega_2$ is the energy of the scattered photon and
$\left({{\rm d}\sigma}/{{\rm d}{\rm \Omega}}\right)_{\mbox{\tiny\rm Th}}$ is 
the Thomson differential cross section.
The outcome of a Compton scattering experiment is therefore given by what
is called a Compton profile in a given direction ${\bf q}$ and defined as
\begin{eqnarray}
J(\tilde{p})=\int {\rm d}{\bf p}\,n({\bf p})\delta\left({\bf p}\cdot\hat{\bf q}-
\tilde{p}\right).\label{cp}
\end{eqnarray}
The Compton profile at $\tilde{p}$ is the integral of the momentum 
distribution on a plane perpendicular to the unit vector $\hat{\bf q}$ at a 
distance $\tilde{p}$ from the origin.
The momentum distribution is expressed in terms of the wave function of
the electronic system as
\begin{eqnarray}
n({\bf p})=\frac{N}{V}\int {\rm d}{\bf r}_1\ldots{\rm d}{\bf r}_N
\int {\rm d}{\bf r}'\; & & e^{i\,{\bf p}\cdot{\bf r}'} \;
\Psi^*({\bf r}_1,\ldots,{\bf r}_N)\,\times\nonumber\\
& & \Psi({\bf r}_1+{\bf r}',\ldots,{\bf r}_N),
\label{np}
\end{eqnarray}
where $N$ is the number of electrons and $V$ the volume of the system.

Since Compton experiments can only access the momentum distribution in an
indirect way and there are difficulties 
in handling background or subtracting the core contribution (the impulse 
approximation may not hold for the core electrons~\cite{ia1,ia2}), 
it is important to establish the performance of the technique for a system 
where high resolution experimental data are available and
correlated calculations can be performed.
Being a low-$Z$ material, lithium has been the subject of Compton studies
since the early days of x-ray Compton scattering~\cite{Eisenberger2}.
Recent high-resolution experiments have been conducted on lithium and Fermi 
surface signatures have been investigated either by analyzing first and second 
derivatives of the Compton profiles~\cite{Sakurai} or attempting a 
reconstruction of the momentum distribution~\cite{Schulke2}. 

Despite overall shape similarities, there is a clear discrepancy between 
the measured Compton profiles of lithium and the theoretical profiles computed 
within Kohn-Sham density functional theory in the local density approximation 
(LDA) even when the Lam-Platzman correlation corrections~\cite{LamPlatzman} 
are included.
These discrepancies were attributed to inadequate treatment of correlation 
in the Kohn-Sham single particle picture~\cite{Sakurai} and it was proposed 
that plasmaron (a bound hole-plasmon state) losses could explain the observed 
difference~\cite{Schulke2}.
Two recent theoretical studies have tried to explain this discrepancy but 
reached opposite conclusions.
Kubo calculated the Compton profiles using the occupation numbers obtained 
from a GW calculation and recovered a good agreement with experimental 
results~\cite{Kubo}.
Dungdale and Jarlborg simulated the effect of thermal disorder on the 
Kohn-Sham Compton profiles, found that disorder leads to a delocalization of 
the momentum distribution and
concluded that this effect, combined with the Lam-Platzman correlation 
corrections, accounts for the discrepancy with the experimental Compton 
profiles~\cite{Dugdale}.

In order to determine the correction to the LDA Compton profiles due to 
electron-electron correlation, we perform a fully correlated calculation 
of the momentum distribution of solid lithium within pseudopotential quantum 
Monte Carlo (QMC).
The QMC Compton profiles differ from the experimental data by Sakurai 
{\it et al.}~\cite{Sakurai} and indicate that Lam-Platzman corrections give 
a satisfactory description of electronic correlation in this system even 
though they are isotropic and cannot reproduce the observed directional 
dependence of the QMC corrections to the LDA Compton profiles.
Therefore, the QMC results for bcc lithium differ from the GW 
calculations~\cite{Kubo} and suggest that the discrepancy between 
conventional band theory and experiments originates from other sources: 
temperature effects~\cite{Dugdale} or problems in the interpretation of the 
experimental data~\cite{ia1,ia2} are possible explanations.

Recently, QMC calculations of Compton profiles have also been performed for
a rather different system, silicon~\cite{Louie1}.
Due to the absence of a Fermi surface, the momentum distribution of silicon 
is a smooth function and finite size errors in QMC are easier to correct 
than in a metal.
These studies concluded that, also in silicon, correlations effects are well
described by the Lam-Platzman corrections and do not fully account for the 
discrepancy between the theroretical Kohn-Sham LDA profiles and the 
experimental results.

%
In Section~\ref{s2}, the characteristics of our density functional
theory calculations are outlined.
In Section~\ref{s3}, the functional form of the QMC wave function is 
described.
In Section~\ref{s4}, we compare the Kohn-Sham Compton profiles obtained 
from pseudopotential and full-core LDA calculations. We then discuss the 
role of electron-ion and electron-electron correlation on the momentum 
distribution in QMC.
We finally present the QMC correlation corrections to the valence Kohn-Sham 
Compton profiles and compare them with the experimental results.
In Appendix~\ref{a1}, the linear tetrahedron method to obtain the Kohn-Sham
Compton profiles is briefly outlined. 
In Appendix~\ref{a2}, convergence of valence properties of lithium from 
full-core plane-wave calculations is discussed.
Variance minimization, variational Monte Carlo (VMC) and diffusion Monte Carlo 
(DMC) methods are briefly presented in Appendix~\ref{a3}.

\section{LDA calculations}
\label{s2}

We study bcc lithium at the experimental lattice constant of $6.60$ a.u. 
We carry out pseudopotential and full-core calculations within LDA density 
functional theory in a plane-wave basis.
The pseudopotential calculations provide the single-particle orbitals that
enter in the QMC wave function and the reference momentum distribution for 
the correlation corrections determined within QMC.
The full-core calculations of the valence contribution to the Compton 
profiles are carried out to account properly for core-valence orthogonality.
In Appendix~\ref{a1}, we describe how to evaluate the Kohn-Sham momentum 
distribution and construct the Compton profiles using the linear tetrahedron 
method.

In the pseudopotential calculations, we use the Troullier-Martins 
pseudopotentials whose {\it s} and {\it p} components are generated with a 
cutoff radius of 2.4 a.u.~\cite{TM}. The plane-wave cutoff is set to 16 Ry 
and 44 special {\it k}-points~\cite{MP} in the irreducible wedge of the 
zone are used for zone sampling during iteration to self-consistency.
The Compton profiles are generated from 16206 {\it k}-points in the 
irreducible zone unfolded in a sphere of 2 a.u.\ radius
(the mesh spacing is $0.0136$ a.u.).
The profiles are converged with respect to mesh spacing and sphere radius but
appear to be very sensitive to the value of the Fermi energy that must be 
carefully determined as described in Appendix~\ref{a1}.

The valence Compton profiles from a plane-wave full-core calculation are 
obtained with a plane-wave cutoff of 400 Ry and 3311 {\it k}-points in the 
irreducible zone, unfolded into a sphere of 4 a.u.\ radius
(the mesh spacing is $0.0238$ a.u.).
The Compton profiles are in good agreement with the profiles computed with 
the linear augmented plane wave method~\cite{Blaas}.
In Appendix~\ref{a2}, we discuss convergence issues for valence properties
computed from a full-core calculation in a plane-wave basis.

\section{Quantum Monte Carlo Calculations}
\label{s3}

In the quantum Monte Carlo simulations of bcc lithium, we treat lithium as 
a pseudo-ion of charge one plus one valence electron. 
We test the use of both a local and a non-local 
pseudopotential~\cite{PSEUDOPOT}.
To simulate an infinite solid, we model the system as a collection of ions
and electrons in a simulation cell periodically repeated.
An effective computation of the Ewald sums over the images of the potential 
is obtained by optimizing the separation between the long- and short-range 
components of the electron-electron, electron-ion and ion-ion
interactions~\cite{ewald}.
We will consider cubic simulations cells containing 54, 250 and 686 lithium 
atoms.

The wave function used in these calculations is a determinant of 
single-particle orbitals multiplied by a Jastrow factor describing 
electron-electron and electron-ion correlations:
\begin{eqnarray}
\Psi={\rm D}^\uparrow \times {\rm D}^\downarrow\exp\left[
-\sum_{i>j}^N u(r_{ij})+\sum_{i=1}^N\chi({\bf r}_i)\right]\,.
\label{wf0}
\end{eqnarray}
${\rm D}^\uparrow$ and ${\rm D}^\downarrow$ are the Slater determinants of 
single particle orbitals for the up and down electrons respectively, $u$ 
correlates pairs of electrons and $\chi$ is a single-body term.

We consider two forms of single-particle orbitals. 
1) We test the very simple choice:
\begin{eqnarray}
\phi_0({\bf r})=e^{i\,{\bf k}\cdot{\bf r}}.\label{phi_simple}
\end{eqnarray}
2) We compute the single-particle orbitals from a LDA density functional theory 
calculation within a plane-wave basis:
\begin{eqnarray}
\phi_{\rm LDA}({\bf r})=
\sum_{\bf G}c_{{\bf k}+{\bf G}}e^{i({\bf k}+{\bf G})\cdot{\bf r}},
\label{phi_LDA}
\end{eqnarray}
where ${\bf G}$ are the reciprocal lattice vector of the underlying bcc
lattice. We discuss the occupation of the orbitals below.

The electron-electron term in the Jastrow factor is periodic over the cell and
contains no free parameters.
It is derived within the random phase approximation (RPA) and has been 
extensively used for the homogeneous electron gas~\cite{rpa} and solid hydrogen 
calculations~\cite{NatoliH}. 
It describes both the exact short range (cusp condition) and large distance 
(plasmon) behavior. We only impose the antiparallel cusp conditions:
\begin{eqnarray}
u^{\uparrow\downarrow}(r_{ij})=u^{\uparrow\uparrow}(r_{ij})=
u^{\downarrow\downarrow}(r_{ij})=u_{\rm RPA}(r_{ij}).
\end{eqnarray}
The single-body term in the Jastrow factor has the periodicity of the 
underlying lattice, is expanded in Fourier components and rewritten as the 
sum over stars of reciprocal lattice vectors:
\begin{eqnarray}
\chi({\bf r})=\sum_s\chi_s \sum_{{\bf G}\in s}
\Phi_{\bf G} {\rm e}^{i{\bf G}\cdot{\bf r}}.
\end{eqnarray}
Since the point group of bcc lithium is symmorphic, the phases $\Phi_{\bf G}$
can be set to unity. We include up to twelve stars but convergence is obtained 
already with an expansion over seven stars. The coefficients of the stars,
$\chi_s$, are optimized using the variance minimization 
method~\cite{optimization,Coldwell}.

A brief description of variance minimization, VMC and DMC methods is
given in Appendix~\ref{a3}.

\subsection{$k$-point sampling}

We impose that the wave function satisfies boundary conditions that 
can be either periodic or arbitrary:
\begin{eqnarray}
\Psi({\bf r}_1,.\,.\,,{\bf r}_i+{\bf R}_s,.\,.\,,{\bf r}_N)
=& &\nonumber\\
e^{i\,{\bf k}_s\cdot {\bf R}_s}& &
\Psi({\bf r}_1,.\,.\,,{\bf r}_i,.\,.\,,{\bf r}_N),\label{bc}
\end{eqnarray}
where one electron has been displaced by a translational vector of the 
simulation cell, ${\bf R}_s$.
The Jastrow component is periodic over the cell and does not affect the
phase of the wave function. 
The determinant satisfies the above equation if the single-particle 
orbitals, $\phi$, obey similar equations, $\phi({\bf r}+{\bf R}_s)=
\exp{\{i\,{\bf k}_s\cdot{\bf R}_s\}}\,\phi({\bf r})$, that is if
the orbitals correspond to a single $k$-point sampling of the 
simulation cell given by ${\bf k}_s$.
If we impose periodic boundary conditions on our cubic cell of side $L$,
the orbitals are Bloch states with wave vectors defined on a cubic grid 
centered on the origin with spacing $2\pi/L$. 
For arbitrary boundary conditions, the grid of wave vectors is shifted by 
${\bf k}_s$.

The idea of $k$-point sampling in QMC was introduced to achieve a faster 
convergence in the total energy versus system size, possibly faster than in 
a periodic calculation~\cite{ksampling}.
For bcc lithium, since the valence band is an $s$-band, a calculation not
at the $\Gamma$-point yields a higher energy and corresponds to an excited 
state of our finite simulation cell. 
On the other hand, the allowed momenta coincide with the wave vectors 
compatible with the condition of periodicity on the simulation cell, so we 
can obtain a higher resolution in momentum space by performing calculations 
with different boundary conditions~\cite{Louie1}.

We only consider $k$-samplings that yield a grid of $k$-points with inversion 
symmetry so that we can construct a real wave function by occupying linear 
combinations of pairs of orbitals:
\begin{eqnarray}
\phi^+_{\bf k}=\frac{1}{2}(\phi_{\bf k}+\phi_{\bf -k}),\,\,\,\,\,\,\,
\phi^-_{\bf k}=\frac{1}{2i}(\phi_{\bf k}-\phi_{\bf -k}).
\label{real_orb}
\end{eqnarray}
By restricting ourselves to real wave functions, we can perform both 
VMC and DMC calculations and avoid the complications of dealing with complex 
wave functions in DMC~\cite{Ortiz}.
For a cubic simulation cell, there are only four vectors ${\bf k}_s$ that 
preserve inversion symmetry corresponding to $\Gamma$, $X$, $M$ and $R$ 
sampling of the cubic simulation cell.
The $\Gamma$-point calculation yields a cubic grid of $k$-vectors centered 
on the origin with spacing $2\pi/L$ and the additional calculations provide 
three grids shifted by $(2\pi/L)/2$ in the [100], [110] and [111] directions 
respectively.

To construct the determinantal part of the wave function, we compute the 
orbitals on the grid of $k$-vectors compatible with the boundary conditions 
and occupy the orbitals, $\phi_0$ or $\phi_{\rm LDA}$, within the first 
Brillouin zone with the lowest free-electron ($\phi_0$) or LDA 
($\phi_{\rm LDA}$) energy. 
In general, the highest occupied level is degenerate and only partially 
occupied, so we should employ a linear combination of determinants in order 
to construct a wave function with the proper symmetry. 
We instead always use a single determinant and symmetrize the momentum 
distribution on the grid of allowed wave vectors by averaging it over symmetry 
related $k$-vectors on the grid. This procedure is done separately for the
four $k$-point samplings. 
Finally, by combining the results of the four calculations and applying
the symmetry rotations of the cubic group, we obtain a momentum distribution 
defined on a mesh with spacing $(2\pi/L)/2$.

We consider cubic cells with 54, 250 and 686 atoms.
For the 54 atom cell, we only carry out a $\Gamma$-point calculation. 
The wave function is constructed from the orbitals corresponding to the lowest 
four complete shells of $k$-vectors, so it is real, has the full symmetry of 
the lattice and is a spin-singlet.
For the 250 and 686 atom cell, we compute the momentum distribution for 
four wave functions corresponding to $\Gamma$, $X$, $M$ and $R$ sampling of 
the cubic cell. 
The grid in momentum space has spacing $0.095$ a.u.\ and $0.068$ a.u.\ for 
the 250 and the 686 atom cell respectively. 

\subsection{Momentum distribution and Compton profiles}

The momentum distribution of the correlated wave function is computed as the 
expectation value over the distribution given either by $\Psi^2$ (VMC) or by 
the product of the trial wave function and the fixed-node solution (DMC):
\begin{eqnarray}
& &n({\bf p})= \nonumber\\
& &\sum_i\frac{1}{V} \left< 
\int {\rm d}{\bf r}'\;e^{i\,{\bf p}\cdot{\bf r}'}
\frac{\Psi({\bf r}_1,.\,.\,,{\bf r}_i+{\bf r}',.\,.\,,{\bf r}_N)}
{\Psi({\bf r}_1,\ldots,{\bf r}_N)}\right>.\label{npqmc}
\end{eqnarray}
At a given Monte Carlo step, we uniformly sample $M$ random positions, 
${\bf r}'$, within the simulation cell and, for each position and particle, 
compute in turn the above ratio~\cite{McMillan}.
The value of $M$ depends on the size of the system and is 
determined to optimize the efficiency in sampling the momentum distribution.  
The allowed values of momentum coincide with the sets of vectors compatible 
with the condition of periodicity imposed on the simulation cell 
(Eq.~\ref{bc}).

To obtain the Compton profiles, we have to integrate the momentum 
distribution over planes perpendicular to a given direction. It is possible
to directly compute the Compton profiles within QMC.
For instance, to evaluate the profile in the [100] direction, we could
estimate the following expectation value:
\begin{eqnarray}
J(\tilde{p}_x)=& & \sum_i\frac{(2\pi)^2}{V} \,\times\nonumber\\
& &\left<\int {\rm d}x\, e^{i\,\tilde{p}_x\,x}\frac{\Psi({\bf r}_1,.\,.\,,
{\bf r}_i+x,.\,.\,,{\bf r}_N)}{\Psi({\bf r}_1,\ldots,{\bf r}_N)}\right>.
\end{eqnarray}
This procedure is equivalent to evaluating the integral of the momentum 
distribution as the histogram over the momenta compatible with the boundary
conditions:
\begin{eqnarray}
J(\tilde{p}_x)\approx({2\pi}/{L})^2\sum_{{\bf p}: p_x=\tilde{p}_x} n({\bf p}),
\end{eqnarray}
and it is clearly a poor representation of the integral of a function with 
several discontinuities, especially if the grid of $k$-vectors is coarse. 
The Compton profiles computed in this way show indeed a strong dependence 
on the size of the simulation cell.

To obtain Compton profiles with reduced finite size errors, we want to
integrate a smoother function than the QMC momentum distribution,
$n_{\rm QMC}({\bf p})$.
We have to select a reference distribution, $n_{\rm MODEL}({\bf p})$, 
whose Compton profiles, $J_{\rm MODEL}( p)$, can be computed with high
accuracy and such that the difference, 
$\Delta n({\bf p})=n_{\rm QMC}({\bf p})-n_{\rm MODEL}({\bf p})$,
is a smoother function than the original $n_{\rm QMC}({\bf p})$.
The Compton profiles are then obtained as
\begin{eqnarray}
J_{\rm QMC}( p)=J_{\rm MODEL}( p)+\Delta J(p)
\end{eqnarray}
where the corrections, $\Delta J(p)$, is computed by integrating 
$\Delta n({\bf p})$ on the grid defined by the four $k$-point samplings 
using the linear tetrahedron method. 
The only difference in computing $\Delta J(p)$ from the method described 
in Appendix~\ref{a1} is that 
there is no Kohn-Sham energy defined on the grid and no Fermi energy but all 
grid points are considered for integration. 
In Section~\ref{s4}, we will show our choice for the reference momentum 
distribution.

\section{Results and discussion}
\label{s4}

We present calculations of the momentum distribution and Compton
profiles within LDA Kohn-Sham density functional theory. We perform both 
pseudopotential and full-core calculations. 
We then compute the momentum distribution within pseudopotential QMC and 
determine the correlation contribution to the directional Compton profiles.

\subsection{Kohn-Sham LDA Compton profiles}

In Fig.~\ref{lda}, we show the LDA momentum distribution along the [110]
and [111] directions for a local ($s$-component) and a non-local 
($s$- and $p$-components) pseudopotential. The effect of the lattice on the 
momentum distribution is more pronounced when using a non-local 
pseudopotential, yielding a bigger reduction below the Fermi break and a
more significant contribution from the secondary Fermi surfaces (unklapp
processes).
For both the local and non-local pseudopotentials, the momentum distribution 
is strongly anisotropic and goes to zero after the first Fermi break since the 
Fermi surface of lithium is completely contained within the first Brillouin 
zone.

\begin{minipage}{3.375in}
\begin{figure}[hbt]
\epsfxsize=9 cm
\centerline{\epsfbox{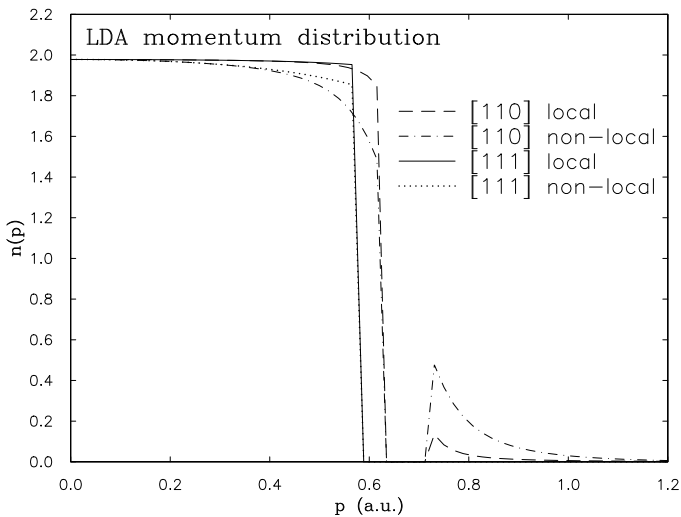}}
\vspace{.2cm}
\caption[]{LDA momentum distribution in the [110] and [111] directions
using a local and a non-local pseudopotential.}
\label{lda}
\end{figure}
\vspace{.7cm}
\end{minipage}

We find that the momentum distribution obtained with the non-local 
pseudopotential is in better qualitative agreement with our full-core 
calculations.
Therefore, for a more realistic description of bcc lithium, we have 
to use a non-local pseudopotential even though this is computationally more
demanding in QMC and introduces the additional locality approximation 
in DMC~\cite{PSEUDOPOT}.

The agreement of the valence momentum distributions from 
a non-local pseudopotential and a full-core calculation is only qualitative
because of the pseudopotential approximation
(lack of correct orthogonalization between core and valence).
When computing the momentum distribution within a pseudopotential scheme, 
we are underestimating the momentum distribution at high momenta and, 
consequently, overestimating it at low momenta. This effect is also
evident in the Compton profiles.
  
In Fig~\ref{fc_pp}, the pseudopotential Compton profiles are compared with 
the valence profiles obtained with the full-core potential in the [100], [110]
and [111] directions. We normalize the Compton profiles to the number of 
valence electrons per unit cell, in this case one electron.
As expected, the Compton profiles constructed from valence orbitals correctly
orthogonalized to the core orbitals show a higher tail at high momenta than
the pseudopotential profiles and, since they must integrate to the 
same value, a significantly lower value at low momenta.
The Compton profile of a free particle system at the same density of
lithium ($r_s=3.25$ a.u.) is an up-side-down parabola terminating at 
$p_{\mbox{\tiny\rm F}}=0.5905$ a.u. 
Due to electron-ion correlation, the Kohn-Sham profiles develop a tail beyond 
$p_{\mbox{\tiny\rm F}}$ but the sharp features of the presence of a Fermi 
surface is clearly visible in any direction at about $0.6$ a.u. 
Its location varies however for the different directions as a result of 
the already observed anisotropy (see Fig.~\ref{lda}).
The effect of secondary Fermi surfaces (unklapp processes) appears both in
the pseudopotential and full-core potential valence profiles. The effect is
more pronounced in the [110] direction at about $0.8$ a.u. and all profiles
show an additional bump in the far tail (above 1.6 a.u.).

\begin{minipage}{3.375in}
\begin{figure}[htb]
\epsfxsize=9 cm
\centerline{\epsfbox{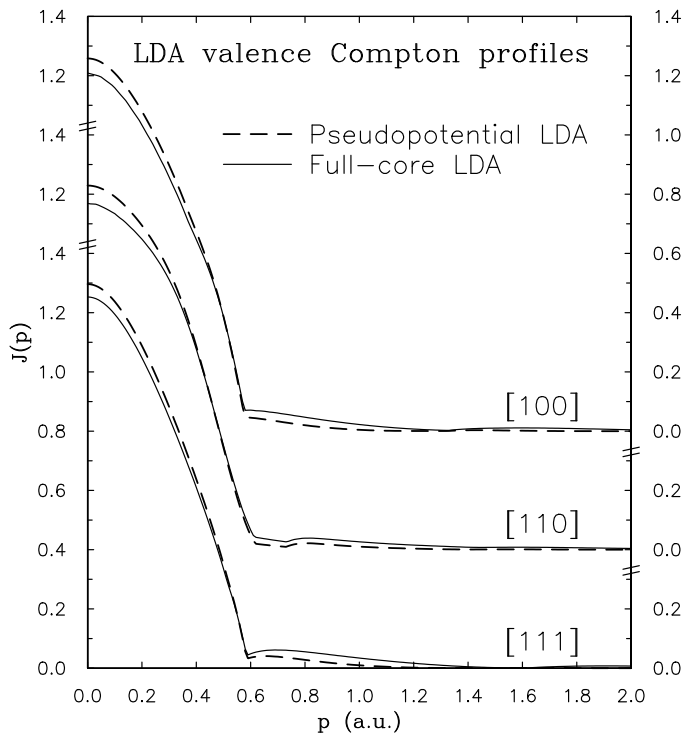}}
\vspace{.2cm}
\caption[]{Valence Compton profiles of Li in the [100], [110] and [111] 
directions.  The LDA Compton profiles constructed using the Troullier-Martins 
pseudopotential are compared with the LDA profiles obtained with the full-core
$-3/r$ potential.}
\label{fc_pp}
\vspace{.7cm}
\end{figure}
\end{minipage}

Within pseudopotential quantum Monte Carlo, we will compute a 
correction to the Compton profiles due to electronic correlations. This
correction will be summed to the valence Compton profile from a full-core
calculation to account properly for core-valence orthogonality.

\subsection{Quantum Monte Carlo Compton profiles}

In order to calculate the correlation corrections to the Kohn-Sham Compton 
profiles of Fig.~\ref{fc_pp},
we first determine potential and wave function needed to correctly describe 
the momentum distribution of solid lithium within QMC. 
We start modeling bcc lithium with the simplest potential and 
wave function and improve upon it with a more sophisticated wave function
and Hamiltonian.

The tests are conducted on a small simulation cell of 54 electrons and 54 
lithium ions on a bcc lattice and with periodic boundary conditions.
%
We carry out QMC calculations using the Troullier-Martins pseudopotential and 
test the use of a local ($s$-component) and a non-local ($s$- and 
$p$-components) pseudopotential as we have done within LDA density functional
theory.
For either potential, we employ both the free-electron orbitals $\phi_0$ 
(Eq.~\ref{phi_simple}) and LDA orbitals (Eq.~\ref{phi_LDA}) determined with 
a plane-wave cutoff of 16 Ry.
The Jastrow factor is separately optimized in each case using 2000 
configurations within variance minimization.
The VMC and DMC energies obtained with the non-local pseudopotential and 
the LDA orbitals are given by E$_{\rm VMC}=-0.2524(1)$ Hartree and 
E$_{\rm DMC}=-0.2591(1)$ Hartree.

In Fig~\ref{pw_lda}, we show the spherical average of the momentum 
distribution in the case of local and non-local pseudopotentials. 
The VMC and DMC spherical momentum distribution for bcc lithium 
are compared with the VMC momentum distribution of the homogeneous electron 
gas at the same density ($r_s=3.25$ a.u.). 
The wave function for the electron gas is given by the product of a determinant 
of simple plane waves, $\phi_0$, and a Jastrow factor only containing the 
electron-electron term described in Section~\ref{s3}.
The momentum distribution of the electron gas shows a discontinuity at 
$p_{\mbox{\tiny\rm F}}$. 
It is reduced below $p_{\mbox{\tiny\rm F}}$ and develops a tail at high 
momenta with respect to the non-interacting step-function.
The persistence af a Fermi break and its location are consistent with the 
Fermi liquid behavior of the system.
In the presence of an electron-ion pseudopotential, we have a further 
reduction of the size of the Fermi break due to electron-ion correlation.
In all cases, the VMC and DMC momentum distribution are almost 
indistinguishable indicating that for this system the variational wave 
function is quite close to the fixed-node solution in describing this property.

For both local and non-local pseudopotentials, we test the use of free-electron 
orbitals, $\phi_0$, versus LDA orbitals.
The use of orbitals given by a single plane-wave is justified if the 
dependence of the periodic component of the Bloch states is not strongly 
varying with ${\bf k}$. 
This dependence can then be simply included in the Jastrow part as 
an electron-ion term, $\chi$ (Eq.~\ref{wf0}). 
As shown in Fig.~\ref{pw_lda}, this argument holds in the case of a local 
potential: the momentum distribution is not significantly different when 
using the orbitals $\phi_0$ or the LDA orbitals.
\begin{minipage}{3.375in}
\begin{figure}[htb]
\epsfxsize=9 cm
\centerline{\epsfbox{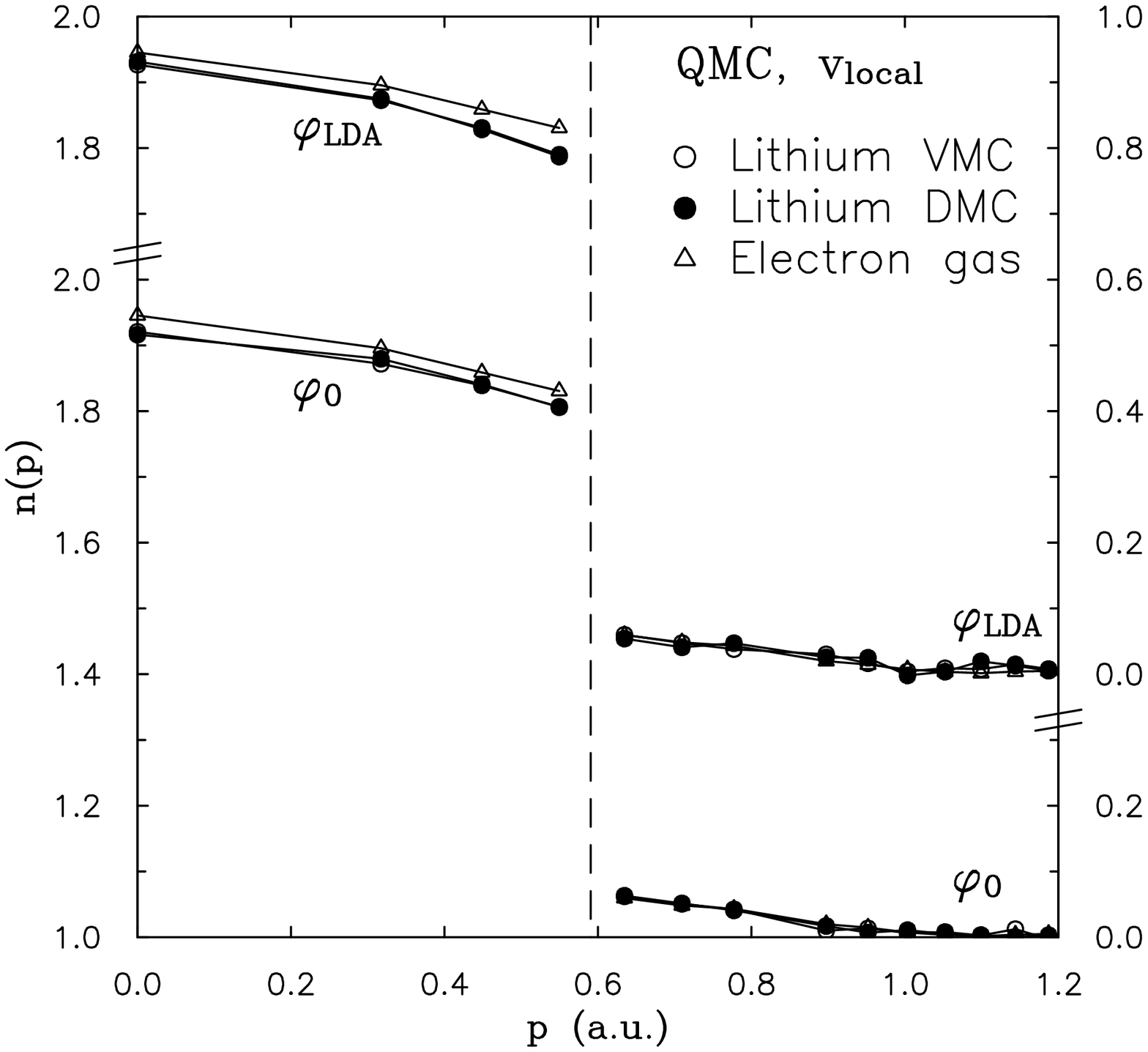}}
\vspace{.2cm}
\epsfxsize=9 cm
\centerline{\epsfbox{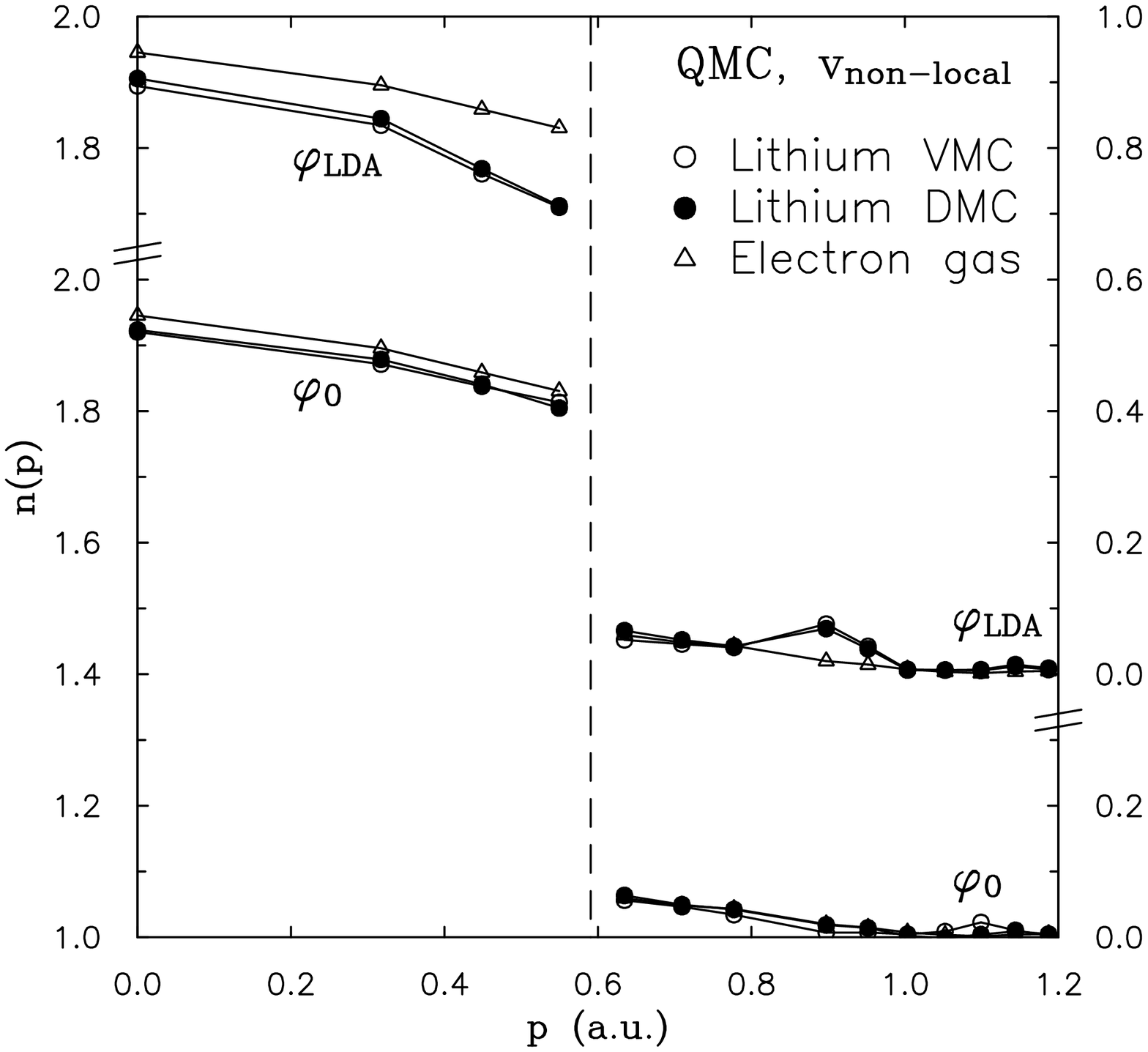}}
\vspace{.2cm}
\caption[]{VMC and DMC spherical momentum distribution for bcc lithium with 
a local (upper plot) and non-local (lower plot) pseudopotential and a 
simulation cell of 54 atoms with periodic boundary conditions.
Either simple plane-wave ($\phi_0$) or LDA orbitals ($\phi_{\rm LDA}$) are 
employed in the determinantal part of the wave function.
The VMC momentum distribution for the electron gas is also shown. 
The statistical errors are smaller than the size of the symbols.
The vertical dashed line indicates the position of $p_{\rm F}$.}
\label{pw_lda}
\vspace{.7cm}
\end{figure}
\end{minipage}
On the other hand, free-electron orbitals in the determinant plus
an electron-ion term in the Jastrow component give a poor description 
of electron-ion correlation when using a non-local pseudopotential.
LDA orbitals yield a larger reduction below the Fermi wave vector and the 
appearance of additional structure in the tail of the momentum distribution 
as contribution from secondary Fermi surfaces.
This effect is smeared out because of spherical averaging and is more 
evident when plotting the momentum distribution along different directions.

Having identified the necessary features to describe bcc lithium (a
non-local pseudopotential and LDA orbitals in the determinatal part of the 
wave function), we perform QMC calculations for larger systems and different 
boundary conditions to obtain a higher resolution in momentum space.
%
We consider two systems with 250 and 686 atoms and compute the VMC momentum 
distribution for four wave functions corresponding to $\Gamma$, $X$, 
$M$ and $R$ sampling of the cubic cell. 
For the system with 250 atoms, we also determine the momentum distribution 
within DMC.
The parameters used in the electron-ion Jastrow component of the four wave 
functions for both system sizes are the ones optimized for the 54 atom 
simulation cell.

As already explained, each $k$-point sampling defines a grid of allowed  
wave vectors and the QMC wave function is constructed from the orbitals with 
the lowest LDA energy. 
This procedure yields a different LDA Fermi energy, 
$\epsilon_{\rm F}^{{\bf k}_s}$, and set of occupation numbers, 
$f^{{\bf k}_s}_{\bf p}$, for each $k$-point sampling. 
The occupation $f^{{\bf k}_s}_{\bf p}$ is equal to two below 
$\epsilon_{\rm F}^{{\bf k}_s}$ and in general fractional at the Fermi level. 
In the following, when comparing the LDA and QMC momentum distribution on 
the grid, we are referring to 
\begin{eqnarray}
n_{\rm LDA}({\bf p})=|\,c_{\bf p}\,|^2\,f^{{\bf k}_s}_{\bf p}\,\,
\theta(\epsilon({\bf p})-\epsilon_{\rm F}^{{\bf k}_s}),\label{LDA_fsize}
\end{eqnarray}
where $c_{\bf p}$ is the Fourier component of an LDA orbital and the momentum 
${\bf p}$ is on the grid corresponding to ${\bf k}_s$-sampling.
This ensures that both the LDA and the QMC momentum distribution satisfy 
the sum sule $\sum_i n({\bf p}_i)=N$ for each $k$-point sampling.

In Fig.~\ref{qmc_den}, we plot the VMC and DMC momentum distributions in the 
[100], [110] and [111] directions for the 250 atom cell.
The LDA distribution evaluated on the same $k$-vectors is also shown for 
comparison.
The VMC and DMC results are quite close to each other with the DMC momentum 
distribution being slightly higher at low momenta and lower at high momenta 
than the VMC one. In the QMC distribution,
we observe a reduction of the LDA momentum distribution below the Fermi wave 
vector, the persistence of the discontinuity and an enhancement at high momenta 
due to electron-electron correlation. 
The momentum distribution for lithium both in LDA and in QMC is strongly
anisotropic. 

\begin{minipage}{3.375in}
\begin{figure}[htb]
\epsfxsize=9 cm
\centerline{\epsfbox{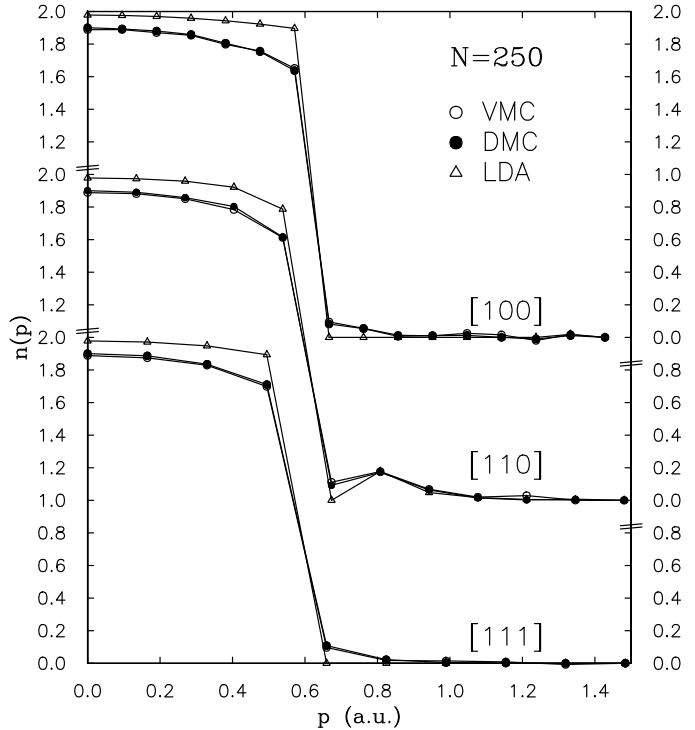}}
\vspace{.2cm}
\caption[]{VMC and DMC momentum distribution of bcc lithium in the 
[100], [110] and [111] directions. The statistical error is smaller than
the size of the symbols.
The momentum distribution is computed for a 250 atom cell and four different 
$k$-point samplings. The LDA momentum distribution evaluated at the same 
vectors is shown for comparison.}
\label{qmc_den}
\vspace{.7cm}
\end{figure}
\end{minipage}

As explained in Section~\ref{s3}, we want to define a reference momentum 
distribution such that its difference with the QMC distribution is a smoother 
function to integrate and Compton profiles with reduced finite size errors
can be obtained.
In Fig.~\ref{al_den}, we show the difference of the QMC and LDA momentum 
distributions along the [110] direction for the 250 atom cell. 
This difference has smaller discontinuity than the original VMC or DMC data 
and its integral is less sensitive to finite size errors. 
However, a yet better choice is to compute the following difference:
\begin{eqnarray}
\Delta\,n^\alpha({\bf p})=n_{\rm QMC}({\bf p})-\alpha\, n_{\rm LDA}({\bf p}),
\end{eqnarray}
where the parameter $\alpha$ is choosen to reduce the size of the 
discontinuity in the function $\Delta\,n^\alpha$. To determine $\alpha$, we 
minimize the cost function:
\begin{eqnarray}
\sum_{{\bf p}}
\sum_{i=1}^3\left[\,\Delta\,n^\alpha({\bf p})-\Delta\,n^\alpha({\bf p}+
{\rm d}p_i)\right]^2\,n_{\rm QMC}({\bf p})^2,
\end{eqnarray}
where ${\rm d}p_i$ is one grid spacing in the $x$, $y$ and $z$ directions.

For both the 250 and the 686 atom simulation cell, we find the optimal value 
of $\alpha=0.8$. 
In Fig.~\ref{al_den}, $\Delta\,n^\alpha({\bf p})$ is plotted in 
the [110] direction for the 250 atom system.
$\Delta\,n^\alpha({\bf p})$ is a more regular function to integrate with no 
significant discontinuity and a similar behavior is observed in the [100] 
and [111] directions. We show below that this procedure of defining a 
reference momentum distribution is successful since it yields Compton 
profiles that lie on the same curve for the various system sizes
(see Figs.~\ref{al_j}~and~\ref{dal_j}).

\begin{minipage}{3.375in}
\begin{figure}[htb]
\epsfxsize=9 cm
\centerline{\epsfbox{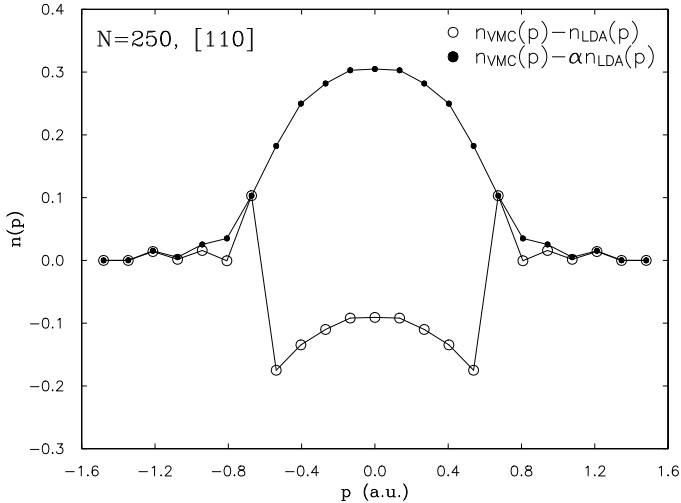}}
\vspace{.2cm}
\caption[]{Difference $n_{\rm VMC}-n_{\rm LDA}$ and $n_{\rm VMC}-\alpha\,
n_{\rm LDA}$ of the VMC and LDA momentum distribution. 
The momentum distribution is computed for a 250 atom cell and four 
different $k$-point samplings. The parameter $\alpha$ is equal to 0.8.}
\label{al_den}
\vspace{.7cm}
\end{figure}
\end{minipage}

To obtain the Compton profiles, we compute $\Delta\,J^\alpha(p)$ by 
integrating $\Delta\,n^\alpha({\bf p})$ with the linear tetrahedron method.
The momentum space is divided in cubes, the cubes in tetrahedra and a 
contribution to the integral from each tetrahedron is computed as in 
Appendix~\ref{a1}.
$\Delta\,J^\alpha(p)$ represents the correlation contribution to the 
Compton profile that is finally obtained as
\begin{eqnarray}
J_{\rm QMC}(p)=\alpha\,J_{\rm LDA}(p)+\Delta\,J^\alpha(p),
\end{eqnarray}
where $J_{\rm LDA}(p)$ is the LDA Compton profile.
To properly take in account core-valence orthogonality, we sum the
correlation contribution $\Delta\,J^\alpha(p)$ to the LDA valence Compton 
profile obtained from a full-core calculation (see Fig.~\ref{fc_pp}).
$\Delta\,J^\alpha(p)$ is computed using a pseudopotential but, since it is 
the difference of results from two pseudopotential calculations (LDA and QMC), 
there is some degree of cancellation in the pseudopotential error. 

In Fig.~\ref{al_j}, the VMC Compton profiles in the [100], 
[110] and [111] directions for the 250 and 686 atom cell are compared
with the LDA valence profiles from the full-core calculation and the 
experimental results by Sakurai {\it et al.}~\cite{Sakurai}.
The agreement between the simulations with 250 and 686 electrons implies that 
the differences between the QMC profiles and either the LDA or the experimental 
curves are beyond finite size errors.
Due to electron-electron interaction, the QMC profiles are lower at low 
momenta and higher at intermediate and high momenta than the LDA profiles.
The experimental data on the other hand show opposite trends. They are lower
than the QMC profiles below the Fermi wave vector and significantly higher 
in the intermediate and far tail. We do not convolute our results with the 
experimental resolution since the momentum resolution for the 250 atom 
simulation is only marginally better than the experimental resolution of 
.12 a.u. 
The continuum line for the QMC profiles is a result of the linear tetrahedron 
construction on $\Delta n^\alpha({\bf p})$. 

\begin{minipage}{3.375in}
\begin{figure}[htb]
\epsfxsize=9 cm
\centerline{\epsfbox{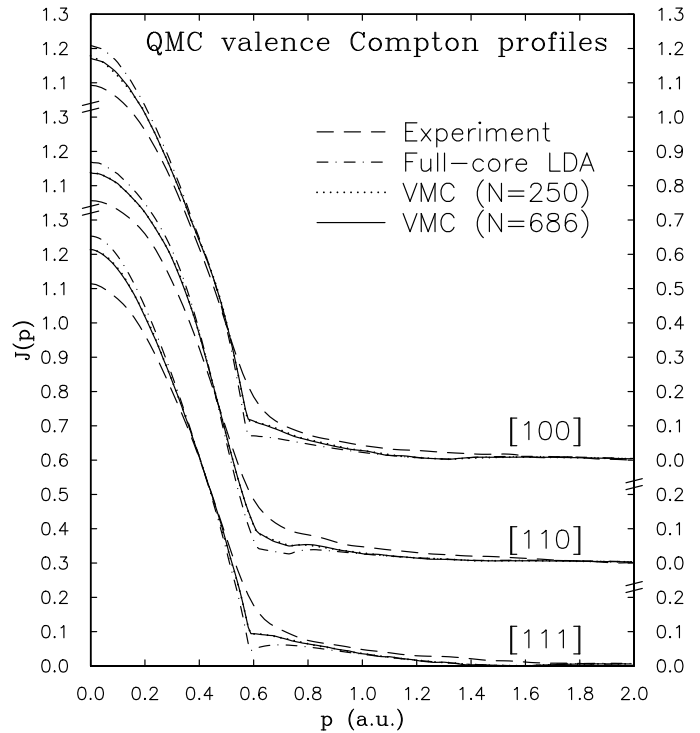}}
\vspace{.2cm}
\caption[]{Valence Compton profiles for lithium in the [100], [110] and 
[111] directions. The VMC results for the cell with 250 and 686 atoms and four 
$k$-point sampling are compared with the LDA valence profiles from a full-core 
calculation and the experimental results~\cite{Sakurai}.}
\label{al_j}
\vspace{.7cm}
\end{figure}
\end{minipage}

In Fig~\ref{dal_j}, we show the difference of the QMC and LDA valence Compton 
profiles, $J_{\rm QMC}-J_{\rm LDA}$, in the [100], [110] and [111] directions,
so details in the correlation corrections can be better appreciated.
We plot the VMC and DMC results for the 250 atom cell and the VMC results 
for the 686 atom cell. 
The agreement between the results of the simulations with the two system
sizes is quite good with the exception of the [100] direction at low momenta 
as a result of finite size errors for the 250 atom cell. 
As already observed for the momentum distribution, the DMC results are only
slightly higher at low momenta and lower at high momenta than the VMC ones.
The difference of the LDA and QMC profiles is anisotropic. 
In the [100] and [111] directions, the QMC profiles do not show the sharp
features of the LDA profiles at the Fermi break but are more rounded off and,
consequently, their difference is peaked at the Fermi break.

\begin{minipage}{3.375in}
\begin{figure}[htb]
\epsfxsize=9 cm
\centerline{\epsfbox{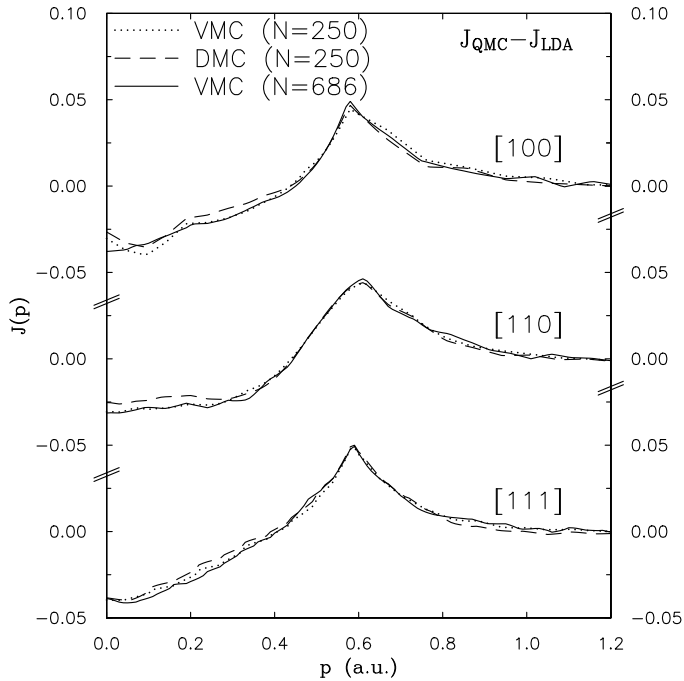}}
\vspace{.2cm}
\caption[]{Difference of the QMC and LDA valence Compton profiles for lithium 
in the [100], [110] and [111] directions. 
We show the VMC and DMC results for the cell with 250 atoms and the VMC results
for the cell with 686 atom.}
\label{dal_j}
\vspace{.7cm}
\end{figure}
\end{minipage}

If we assume a constant valence density corresponding to $r_s=3.25$ a.u.\
and employ the QMC momentum distribution we computed for the electron gas,
the resulting Lam-Platzman corrections are in qualitative agreement with the 
QMC correlation corrections of Fig~\ref{dal_j}. 
Consequently, given the large difference between the experimental results 
by Sakurai {\it et al.} and the QMC profiles, the Lam-Platzman correction 
offer a satisfactory description of electronic correlation in QMC even though 
they are isotropic and cannot resolve the directional differences 
observed within QMC. 

In disagreement with the GW calculations~\cite{Kubo},
the QMC results indicate that electronic correlation in lithium only accounts
for about 30\% of the discrepancy between experimental data and conventional 
band theoretical results. A similar conclusion has been recently obtained
in a comparison of QMC and experimental Compton profiles for bulk 
silicon~\cite{Louie1}. Other effects such as temperature~\cite{Dugdale} or 
the interpretation of experimental results (in particular, the validity of
the impulse approximation~\cite{ia1,ia2}) may explain the 
difference between experiments and QMC calculations.
Our calculations also show that QMC can not only offer a qualitative 
description of correlation effects in Compton profiles but, if finite size 
errors are properly taken care of, also resolve directional differences in 
the correlation corrections to the LDA results and provide useful comparisons 
for new Compton scattering experiments~\cite{HamalainenLi}.

\acknowledgments
This work is supported by NSF (grant DMR 9422496). 
We are grateful to Dr.\ Y.\ Sakurai for sending us the experimental results
and to Prof. Jose Luis Martins for giving us his plane-wave code.
We thank Bernardo Barbiellini and P. Platzman for suggesting this project 
and for many useful discussions.
C.\ F.\ also benefited from several discussions with Erik Koch, Keijo
H\"am\"al\"ainen, Aleksi Soininen and Laurent Bellaiche. 
The calculations were performed at the National Center for Supercomputing
Applications of the University of Illinois at Urbana-Champaign.

\appendix

\section{Linear tetrahedron method}
\label{a1}

To determine the momentum distribution within Kohn-Sham density functional 
theory, the Kohn-Sham valence wave functions and eigenvalues are evaluated 
on a very fine grid of $k$-points within the irreducible wedge of the 
Brillouin zone. 
The Fourier coefficients of the orbitals with the corresponding eigenvalues 
are unfolded from the irreducible zone to full space and the momentum 
distribution is obtained by simply squaring the Fourier components.
This procedure yields a square mesh with the momentum distribution and 
single-particle energy defined at each point. 

To obtain the Compton profiles in a given direction, we integrate the momentum 
distribution using the linear tetrahedron method as described by Lehmann and
Taut~\cite{tetra}. 
The expression for the Compton profile (Eq.~\ref{cp}) is rewritten as
\begin{eqnarray}
J(\tilde{p})=\int{\rm d}{\bf p}\,n({\bf p})
\,\delta(\hat{\bf q}\cdot{\bf p}-\tilde{p})
\,\theta(\epsilon({\bf p})-\epsilon_{\mbox{\tiny\rm F}}),
\end{eqnarray}
where $n({\bf p})$ and $\epsilon({\bf p})$ are the Kohn-Sham single-particle 
momentum distribution and energy and $\epsilon_{\mbox{\tiny\rm F}}$ the Fermi 
energy. The $\theta$-function is introduced to ensure that only states
below the Fermi energy are included.
If $n({\bf p})$ and $\epsilon({\bf p})$ have been computed
on a square grid in momentum space, the mesh naturally divides space in cubes 
whose corners are defined by the grid points. Each cube is then divided in six
tetrahedra and each tetrahedron is considered in turn.
Within each tetrahedron, $n({\bf p})$ and $\epsilon({\bf p})$ are linearly 
interpolated. 
The step function $\theta(\epsilon({\bf p})-\epsilon_{\mbox{\tiny\rm F}})$ 
may restrict the integration to only part of the tetrahedron: the primary 
tetrahedron is divided in secondary ones delimited by the boundaries 
$\epsilon({\bf p})=\epsilon_{\mbox{\tiny\rm F}}$.
For a given value of $\tilde{p}$,
the surface of constant projection, $\hat{\bf q}\cdot{\bf p}=\tilde{p}$, is 
determined within each of the secondary tetrahedra and the contribution 
to $J(\tilde{p})$ is computed as the integral over this surface of the linear 
interpolation of the momentum distribution.

Before computing the Compton profiles, the Fermi energy 
$\epsilon_{\mbox{\tiny\rm F}}$ must be estimated. For a given value of the
Fermi energy, we apply the linear tetrahedron construction to determine the 
volume delimited by the Fermi surface 
and iteratively change $\epsilon_{\mbox{\tiny\rm F}}$ so that 
\begin{eqnarray}
\int{\rm d}{\bf p}\,\theta(\epsilon({\bf p})-\epsilon_{\mbox{\tiny\rm F}})=
\frac{4\pi}{3}p_{\mbox{\tiny\rm F}}^3,
\end{eqnarray}
where $p_{\mbox{\tiny\rm F}}$ is the Fermi momentum.

\section{Full-core plane-wave LDA calculations}
\label{a2}

As mentioned in Section~\ref{s4},
due to the lack of correct orthogonalization between core and valence,
the pseudopotential momentum distribution is too low at high momenta and,
consequently, too high at low momenta. 
To account for the correct oscillatory behavior of the valence wave functions, 
the valence orbitals are usually computed within traditional all-electron 
schemes (LAPW or LMTO) or, alternatively, the true valence wave functions 
are reconstructed from the pseudized orbitals~\cite{Meyer,Louie2}.
On the other hand, several calculations in the literature have shown the 
feasibility of plane-wave calculations using the unscreened Coulomb potential 
for first-row elements~\cite{NatoliH,Bellaiche,Teter}.
We therefore decided to follow this more straightforward route and adopt a 
plane-wave basis also with the $-3/r$ potential of lithium.

In table~\ref{t1}, we show the convergence in the total energy and the
$1s$ and $2s$ eigenvalues at the $\Gamma$ point for full-core solid lithium 
as a function of plane-wave cutoff. For zone sampling to selfconsistency, we
used 14 special $k$-points. To obtain an accuracy of a few mRy on either the 
total energy or the 1{\it s} eigenvalue, plane-wave cutoffs higher than 1800 Ry 
are required. On the other hand, the convergence in valence properties 
such as the 2{\it s} level is achieved at significantly lower cutoffs. 

Bellaiche and Kunc~\cite{Bellaiche} obtained convergence in the structural 
properties of solid LiH with a plane-wave cutoff of the order of 200 Ry and 
the valence Compton profiles were in good agreement with the ones reconstructed 
by simply orthogonalizing the valence wave functions to a core atomic orbital.
Similarly, we find that the valence orbitals of lithium are well converged at 
about 300 Ry and increasing the plane-wave cutoff to 400 Ry has a negligible 
effect on the valence Compton profiles.

\begin{minipage}{3.375in}
\begin{table}[phbt]
\caption[]{Total energy of full-core bcc lithium versus the energy 
cutoff employed in the plane-wave calculation. The energy of full-core atomic 
lithium is $E_{\rm atom} =-14.6682$ Ry. All energies are in Rydberg.}
\label{t1}
\begin{tabular}{lccc}
$E_{\rm cutoff}$ & $E_{\rm total}$ & $\epsilon_{1s}$ & $\epsilon_{2s}$\\[.05cm]
\hline\\[-.3cm]
 300  & -14.6277 & -2.7442 & -0.3102 \\ 
 400  & -14.6886 & -2.7619 & -0.3086 \\
1500  & -14.7985 & -2.7948 & -0.3058 \\
1800  & -14.8032 & -2.7962 & -0.3057 \\
\end{tabular}
\end{table}
\end{minipage}

\section{Quantum Monte Carlo Methods}
\label{a3}

The variance minimization method~\cite{optimization,Coldwell} consists of 
the minimization of the variance of the local energy
over a set of $N_c$ configurations $\{R_i\}$ sampled from the square of
the best wave function available before we start the optimization, $\Psi_0$:
\begin{eqnarray}
\sigma_{\rm opt}^2[\Psi]=\sum_i^{N_c}\left[
\frac{{\cal H}\Psi(R_i)}{\Psi(R_i)}-{\rm E}_{\rm guess}\right]^2 w(R_i)/
\sum_i^{N_c} w(R_i).
\end{eqnarray}
E$_{\rm guess}$ is a guess for the energy of the state we are interested in
and $w(R_i)=\left|\Psi(R_i)/\Psi_0(R_i)\right|^2$.
We do not allow the ratio of the weights to the average weight to exceed a
maximum value.

We compute the expectation value of various operators both in variational
and diffusion Monte Carlo.
In VMC, configurations are sampled from $\Psi^2$ using Metropolis Monte
Carlo method and the expectation value of a given operator ${\cal O}$ is 
obtained from
\begin{eqnarray}
{\rm O}_{\rm VMC} = {1 \over N}\sum^N_i {{\cal O} \Psi(R_i) \over \Psi(R_i)}.
\end{eqnarray}
The transition matrix consists of a drift-diffusion step with a time-step
optimized to minimize the auto-correlation time.
In DMC, the imaginary-time evolution operator $\exp(-{\cal H}\tau)$ is used 
to project out the ground state from the trial wave function within the 
fixed-node and the short-time approximations~\cite{dmc}.  
The time-step error coming from the short-time approximation is negligible 
but the fixed-node error limits the accuracy of the results we obtain.

\end{multicols}

\end{document}